# A Quantum Cellular Automata architecture with nearest neighbor interactions using one quantum gate type.

Dimitris Ntalaperas and Nikos Konofaos

Department of Informatics

Aristotle University of Thessaloniki,

Biology Building, Main University Campus, 54124 Thessaloniki, Greece

Abstract

We propose an architecture based on Quantum cellular Automata which allows the use of only one type of quantum gates per computational step in order to perform nearest neighbor interactions. The model is built in partial steps, each one of them analyzed using nearest neighbor interactions, starting with single qubit operations and continuing with two qubit ones. The effectiveness of the model is tested and valuated by developing a quantum circuit implementing the Quantum Fourier Transform. The important outcome of this validation was that the operations are performed in a local and controlled manner thus reducing the error rate of each computational step.

# 1 Introduction

The notion of taking advantage of quantum mechanical principles in order to design computers that could, in principle, be more powerful than their classical counterparts was firstly proposed by Feynman [1], while Margolus [2] demonstrated the existence of computational models that were based on quantum mechanics and could perform reversible computation in the form of Reversible Cellular Automata (RCA). Various quantum computational models, being able to exhibit universal behavior, have also been defined and shown to be equivalent. In particular, Deutsch [3] defined the formulation of the Quantum Turing Machine (QTM), Yao [4] formulated the Quantum Circuit Model and demonstrated its equivalence with the QTM, while Watrous [5] defined one-dimensional Quantum Cellular Automata and showed that there is an efficient way to simulate a subclass of them by a QTM.

Algorithms that harness the power of quantum computational model and outperform the corresponding classical ones have also been developed; Shor [6] has designed a quantum algorithm that can factorize composite numbers achieving a superpolynomial speedup, while Grover [7] demonstrated how searching in an unsorted database can be performed by a quantum algorithm with a quadratic speedup.

Physical implementation of a quantum computer can pose a challenge, since quantum systems tend to decohere due to unwanted couplings with the environment. Various implementation architectures that try to overcome this problem have however been proposed [8] [9].

One common characteristics of the vast majority of implementation architectures is that they allow for only neighboring qubits to interact. Due to this limitation a number of schemes have been studied, in which a general quantum algorithm can be converted to an equivalent one, allowing only nearest neighbor interactions [10][11]. Typically, these schemes convert a quantum circuit to one consisting of gates acting only on neighboring qubits at the cost of introducing a number of additional gates to the original circuit.

In this paper, an architecture that enforces only nearest neighbor interactions and the application of only one type of a quantum gate per computational step is introduced. This new architecture is based on Quantum Cellular Automata (QCA). The goals of the new architecture are: a) To convert a generic quantum circuit to an equivalent one having only nearest neighbor interactions and b) to allow only a specific quantum gate to act on all the qubits for the duration of one computational step. This latter limitation is introduced since in many implementations it is difficult to localize signals corresponding to two different quantum gates. In the architecture presented, a single signal is applied in each step and the same error is introduced to all qubit states. The only non local interactions allowed are those of quantum teleportation which have been shown to have a nearly zero error rate in various implementation schemes [12].

The present paper is structured as follows: First, an overview of a QCA is given along with an overview of specific type of QCA proposed by Karafyllidis [13] , which operate by applying a single two-qubit quantum gate over the whole quantum register in each step. Afterwards, an overview of nearest neighbor architectures is presented and it is demonstrated how quantum gates involving qubits over an arbitrary distance can be performed by using only local interactions and quantum teleportation in a 2D Grid model architecture which is based in ideas introduced by Rosenbaum[14] Then, we introduce our proposal where the two techniques are combined to this new architecture that retains the advantages of both methods. The execution of an arbitrary quantum gate is demonstrated in this new model and the amount of extra operations required per gate is calculated. Finally, an example of developing a quantum circuit implementing the Quantum Fourier Transform is given, which relies on our model.

## 2 Theory

### 2.1 Cellular Automata

Arrighi et al [15] gave the axiomatic definition of a QCA by defining QCA as a unitary operator *G* over a quantum labeled graph (QLG). Briefly, a QLG is defined as a tuple

$$\Gamma = (V, E, H)$$

With *V* being the nodes, *E* a subset of *VxV* being the edges and *H* a set of Hilbert spaces being the labels. $H^x$ denotes the Hilbert space of node *x* with $\Sigma^x$ its countable canonical basis. Using this definition, a QCA is defined as a unitary operator $G: H \to H$ over QLG $\Gamma$ with the following restrictions:

- *V* forming a grid ($V = Z^n$)
- *E* connecting nodes *x* to *x+z*, with $x \in Z^n$ and $z \in \{0,1\}^n$
- *H* being of finite dimension

Various QCA operational descriptions (e.g. Perez et al [16]) can be shown to be equivalent to the axiomatic description proposed by Arrighi et al. Of convenience, are descriptions that model the unitary operations with a quantum gate, which result in quantum circuit structures for QCA.

One such a quantum circuit structure was proposed by Karafyllidis [13]. In this structure, there are two qubits per cell, the s-qubit (state qubit) and the c-qubit (controlled qubit). At each computational step t the state of the jth cell can be written as $|s_j^t c_j^t\rangle$. Since there four basis cell for each cell, the general state of each cell during evolution is:

$$|s_j^t c_j^t\rangle = c_{0,j}^t |00\rangle + c_{1,j}^t |01\rangle + c_{2,j}^t |10\rangle + c_{3,j}^t |11\rangle$$

with $c_{i,j}^t$ being complex numbers. The state of the QCA consists of the tensor product of all the cell states and can be written as

$$|S\rangle = \otimes_{i=1,n} |s_i^t c_i^t\rangle$$

In each step, a controlled operation between a control and a target qubit of adjacent cells takes place, followed by a unitary transformation of the two qubits of each cell. State $S^{t+1}$ is obtained from state $S^t$ by applying operator R via:

$$|S^{t+1}\rangle = R|S^t\rangle$$

with the Rule R given by:

$$R = R_e R_I$$

with $R_e$ describing the evaluation phase and $R_I$ describing the interaction phase and given respectively by:

$$R_e = \cdots \otimes U \otimes U \otimes \cdots$$

$$R_I = \cdots \otimes CN \otimes CN \otimes \cdots$$

where *U* denotes an arbitrary unitary operation and *CN* a Controlled-NOT operation.

The quantum circuit depicted in Figure 1 performs this implementation. This architecture exhibits a periodic structure in the output probability patterns.

## 2.2   Nearest Neighbor Circuits

Most implementation techniques allow for only local interactions between qubits, therefore prompting the development of various techniques for converting a generic quantum circuit to one consisting only of gates that act upon neighboring qubits These techniques typically make use of the SWAP gate and result in a significant increase of the number of quantum gates of the circuit. Saeedi et al. [11] have developed sophisticated algorithms which can reduce the cost of this conversion compared to the naïve approaches to a factor up to 82%, with the average improvement being around 50%. On the other hand, Rosenbaum [14] has demonstrated a set of techniques for applying a series of reallocations of qubits in a 2 dimensional grid by using quantum teleportation, so that qubits that are to interact over a distance of more than one qubit apart are transported to adjacent positions.

Allowing for teleportation in two dimensions, may further reduce the cost of adding extra gates for achieving nearest neighbor architecture, since the reordering can be done in fewer operations.

Quantum teleportation has been studied and shown to be feasible in various models. D'Ariano et al. [17] have demonstrated the existence of QCA that achieves approximate phase-covariant cloning of qubits while Pfaff et al [12] have demonstrated the teleportation of arbitrary quantum states between diamond spin qubits over long separations. Of particular interest to QCA is the work of Brennen and Williams [18] who have defined optimal pulse sequences for distributing entanglement over a QCA in a 1 dimensional Ising spin chain.

In the following paragraphs, we give a general description of how a general quantum gate between qubits positioned at arbitrary positions can be performed. Since any quantum gate can be decomposed to a set of gates having at maximum two qubits for input, only interactions between two qubits are considered in this work.

Figure 2 depicts the initial configuration of a qubit register.. The register resides at the leftmost part of the two dimensional grid. Same colors indicate that the two qubits will be the input to the same quantum gate. In the particular example, qubit 1 is to interact with qubit 4, qubit 3 with qubit 5 and qubit 2 will be operated by a single qubit operation. Only teleportation in horizontal and vertical directions of the grid should be allowed

The listing in Figure 3 shows the general algorithm for relocating the qubits on the grid. A global *position* variable holds the current depth to which a qubit will be teleported. For each qubit it is examined whether it partakes in a quantum gate operation. Two cases are distinguished:

- The qubit partakes in a single qubit operation. In this case the qubit is teleported horizontally to *position*. *Position* is decreased by one and the qubit is marked as *teleported*. The case where the qubit partakes in no operation at all is handled the same way.
- The qubit partakes in a two qubit operation. In that case the qubit is not marked as teleported, and it is teleported horizontally to *position*. The qubit it interacts with is then teleported to *position-1*. *Position* is decreased by two and the two qubits are marked as *teleported*.

Marking in the above procedure is done classically by marking the index *i* the qubit has in the quantum register by using classical software.

Figure 3 shows a snapshot example of the algorithm when run in the configuration depicted in Figure 2. It is clear that if all the qubits are teleported vertically down to the bottom row and only local interactions are needed in order to execute the computational step required at the starting configuration.

## 3   Arbitrary quantum operations

The two models presented before are now combined in order to develop a new model for quantum gate interactions. The goal of the model is twofold:

- Each operation is to be performed only by neighboring qubits
- In each step of the computation, a single two qubit gate is to be applied to the whole register holding the current state of the system.

Quantum teleportation is not considered in the set of operations mentioned above, since as mentioned, there are various architectures that allow for nearly error free quantum state transportation. Avalle and Serafini [19] have introduced a class of completely positive maps on a qubit array which generalizes the notion of a QCA allowing it to capture all possible stochastic maps on classical probability distributions Their work allows for the modelling and computation of error due to dephasing or amplitude damping over a general quantum channel.

As previously stated, only one qubit operations and controlled two qubit operations are considered. Indeed, there exist such sets of quantum gates (e.g. Hadamard gate, the π/8 gate, and the Controlled NOT gate) that are suitable for universal quantum operations [20].

Since the mechanism for each case (single qubit operations and controlled operations) is different, each case will be treated separately.

### 3.1 Single qubit operations

Suppose a single qubit operation is to be performed on some of the qubits. The nearest neighborhood condition is then automatically fulfilled. If the operation is applied to the whole register however, all the qubits will be altered even those not partaking to the computation. In order to circumvent this, all single qubit operations are converted to controlled operations. The goal is to configure the register so the qubits needed for the computation will be controlled by ancilla qubits which are initially in prepared states |1> and |0>. Figure 5 shows an example of the configuration of a grid; qubits with dashed lines are

those that need to be transformed according to some unitary operation *U*. Grey qubits are ancila qubits prepared in state |0>, while qubits in black color are prepared in state |1>.

In accordance with the procedure described in Section 2.1 the qubits will be teleported to their corresponding positions at the bottom line of the grid. In the next step the ancila qubits are teleported according to whether the operation is to be performed or not. The final state of the grid after all these teleportation procedures is depicted Figure 6. After the state is prepared the controlled-U operation can be applied simultaneously to the whole register consisting of the data and the teleported ancila data. Thus, by applying a signal operation across the whole register the equivalent operation of localizing the operation has been achieved at the cost of 3N extra qubits and the 2N teleportation operations.

## 3.2 Two qubit quantum operations

For two qubit operations the algorithm presented in section 3 is slightly modified. The second qubit (the one used as a control) is teleported to the same column as the first qubit and is teleported at the place where the corresponding ancila qubit of the column resides (Figure 7). The controlled operations are performed in a similar way as the one described in section 3.1, where the qubits not partaking in the operations are used as input to a controlled operation with a control qubit set to state |0> via teleportation. The control qubit can then be teleported to the "vacant" spot (circle in red outline in Figure 7) by using two teleportation operations, so that the data register may once again reside in a single line of the grid. The number of operations is the same as those in 3.1 plus two extra steps for the final teleportation.

## 4 General Algorithm Flowchart

In order to better visualize the execution, the flowchart of the general algorithm is presented in this section which summarizes all the functionalities of the algorithm. The flowchart is depicted in Figure 8. The algorithm loops for *N* iterations, *N* being the size of the extended original quantum circuit. Extended circuit in this context means that quantum gates are performed one at a time, except for the case that the same gate is applied to different places in the quantum register. In the latter case these gates are grouped in the same step in the extended circuit.

The algorithm converts the computational step of each gate to a set of equivalent gates consistent with the rules presented so far.

- If the quantum gate is a single qubit gate the transformation presented in Section 3.1 (Figure 6) is applied
- If the quantum gate is a two qubit gate the transformation presented in Section 3.2 (Figure 7) is applied

This process continues for *N* steps and the final measurement is performed to the working register. As is shown in **Σφάλμα! Το αρχείο προέλευσης της αναφοράς δεν βρέθηκε.**, it is essential to restore any ancila registers to their initial status as they will be used again each time the working register is rotated back to the same position.

# 5  Application - Quantum Fourier Transform

Quantum Fourier Transform resides at the core of Shor's algorithm. The circuit implementing the Quantum Fourier Transform is demonstrated recursively in Figure 9. In each step, qubit *n* is used as a control for controlled phase operations on each of the other qubits. A single Hadamard transformation is applied to qubit *n* at the end.

For the purpose of demonstrating how the proposed architecture works in a generic scenario, we will show the details of how a sequence of controlled phase and a Hadamard gate is performed. This can be extended to the complete QFT circuit, since $QFT_n$ consists of *n-1* controlled phase operations plus a Hadamard operation.

Firstly, we consider the state of the register at some instance with each qubit being in an unknown superposition which depends on the previous computational steps. For the case of a five-qubit register this is depicted in the left side of Figure 10. Each qubit is represented with a different shade of green color. This shade is used to mark the position of the qubit throughout the computation and is not representative of the qubits state.

At first controlled $R_{\pi/2}$ gate is to be performed between qubits at row positions *1* and *2*. In accordance with the algorithm for performing a two-qubit gate, the qubits are first teleported horizontally as displayed in the right side of Figure 10.

The process continues by vertical teleportation and the grid reaches the configuration depicted on the left side of Figure 11. In the same state we demonstrate the teleportation of qubits in state |0> to the relevant positions in the ancila qubits row. The ancila qubit in column *4* is not used and thus the corresponding teleportation is not performed.

After the vertical teleportation the grid is in a configuration that the actual computation may take place. This is shown on the right side of Figure 11. Each box represents a controlled $R_{\pi/2}$ operation. The first four operations leave the computational qubits unaffected since the ancila qubit which is used as a control is in state |0>. The last operation (denoted by the box in red outline in the right side of Figure 11) is the actual $R_{\pi/2}$ operation between qubits *1* and *2* that is required by QFT. All these operations are performed in parallel, so that in each step a single two qubit operation is performed to the whole register and the ancilla qubit array. As required by the model the operations are all nearest neighbor only and a single kind of signal is required in order to perform the computation.

In order to complete the computation, the Hadamard gate must then be performed on the first qubit. At first, the cleanup actions of the previous operation must be performed. This, as described in Section 3.2, involves the teleportation of qubit *2* to the "vacant" position in of the register. After the teleportation the register that now containing the original qubits can now be used for initiating the procedure needed for the Hadamard Gate. This initial configuration is depicted in the left side of Figure 12. The grid is depicted "rotated". This rotation is not physical, but rather a visual cue to help visualizing the process; the qubits should be considered to occupy the same places in the grid as before.

In the right side of Figure 12, the configuration of the grid after the first horizontal teleportation may be seen. This step is similar as before, the exception being that now all qubits are teleported in their respective columns since no qubit will act as control.

The algorithm continues with the vertical teleportation as can be seen in the left side of Figure 13. In parallel the ancila qubits needed are teleported. For this operation the ancila qubits of the first four columns are set to state |0> by teleporting qubits already set to state |0>. This ensures that the state of the corresponding qubits in the working register will remain unaltered. For the 5th column, a qubit in state |1> is teleported in the corresponding ancila position. In the same manner as with the previous gate, a controlled Hadamard is performed pair wise to all qubits of the working register and the ancila register. Since only the qubit of the fifth column is affected, this operation is equivalent to a Hadamard gate performed on said qubit.

It has been demonstrated thus that the proposed model is at least capable of performing any computation that can be described by a Quantum Circuit and is in this sense Turing complete.

For computing the total additional space and number of steps that the model introduces compared to a quantum circuit of size *N we* note the following:

- The grid needs $N^2$ working qubits plus *4N* qubits for the row of the ancila qubits. An additional *2N* qubits holding prepared states of |0> and |1> is needed if we wish to perform all the ancila teleportation operations in parallel.

- Each computational step needs *N* horizontal teleportation operations plus *N* vertical (the extra two teleportation in the case of two-qubit gates can be ignored). There are also at max *N* teleportation operations for the ancila qubits plus at max *N* initialization operations since we need the qubits holding the prepared states to be returned to their original states after the teleportation operations. Finally, there are *N* controlled operations in each phase of the computations

All the above give a cost of additional $O(N^2)$ space and $O(N)$ additional steps. However, as has been demonstrated, the additional steps can be performed in parallel with each operation being executed in three phases. Phase one consisting of the initial horizontal teleportation, phase two consisting of the vertical teleportation and phase three consisting of the quantum controlled operations (ancila teleportation operations and re-initialization may be performed during phases one and/or two).

The benefit from all these however, is that the model permits only nearest neighbor quantum operations which can be performed using localized signals. Furthermore, a single kind of signal is performed during each computational step. It is furthermore to be noted that, ignoring quantum teleportations, which can be regarded as qubit transfer operations, the depth of the original circuit does not change despite the introduction of extra quantum gates. This is due to the parallel execution of these gates in each step and it is contrast with various nearest neighbor architectures proposed so far that require an increase in the quantum circuit's depth in the general case. Including teleportations, the increase of circuit depth is up to a constant per each step.

## 5.1 Flowchart of Quantum Fourier Transform Operations

The Quantum Fourier Transform consists of a set of Hadamard and controlled Phase operations. The various stages of the execution of each gate and their correspondence with the new algorithm's flowchart are depicted in Figure 14 (single qubit operations) and Figure 15 (two qubit operations). Each of the two possible sets of operations of the main loop of the algorithm is mapped to the state of the computational grid with arrows denoting teleportation operations. As can be seen, apart from these teleportations, only one type of quantum gate is applied in the last stage of the procedure to the whole working register.

# 6 Conclusions

In this paper a novel architecture has been proposed with the aim of combining the techniques of Quantum Cellular Automata and Nearest Neighbor interactions. The proposed architecture benefits from the advantages of both techniques; QCA model offer simplicity in design and the capability of applying a gate array consisted of similar gates, whereas the Nearest Neighbor model offer an easier implementation scheme and a low error rate introduction.

While the architecture demands that some extra steps and some extra qubits are to be introduced (of linear and quadratic order respectively), this is compensated by the fact that operations are performed in a local and controlled manner, thus reducing the error rate of each computational step.

# 8 Figures

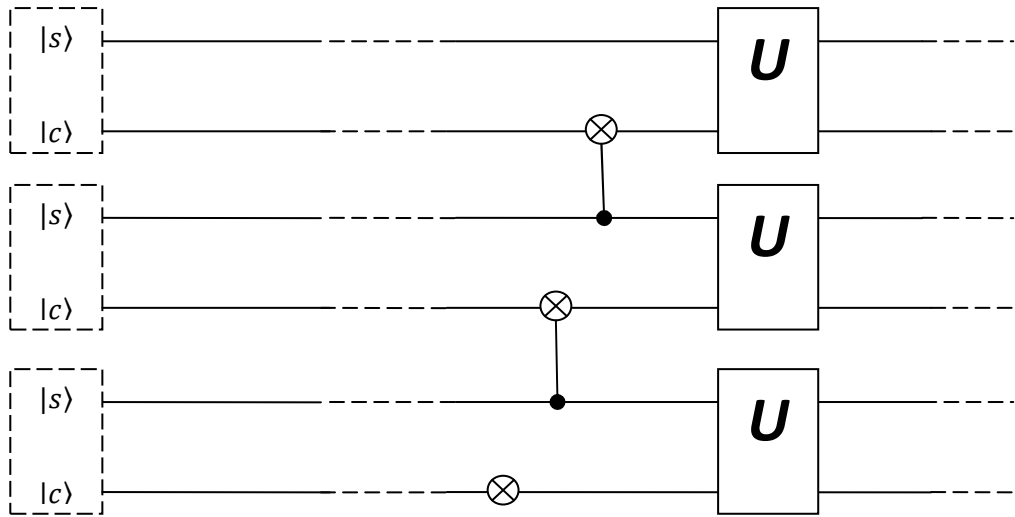

Figure 1 Quantum Cellular Automata architecture, as proposed in reference [6].

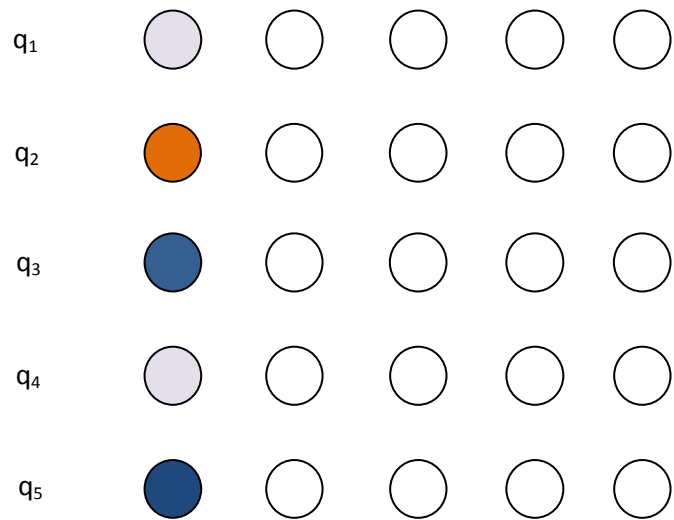

**Figure 2** Nearest Neighbor interactions in a two dimensional grid. Initial configuration. Qubits of the same color are to interact with each other while qubits having a unique color will be acted upon by a single qubit gate.

```
Int Position = N

Bool Teleported[N]

for i 1 to N

    if (qubit(i) interacts_with qubit(j) && !Teleported[j])

        Teleport (qubit(i), Position)

        Teleport (qubit(j), Position-1)

        Position -= 2

        Teleported[i]=Teleported[j]=true

    endif

    if (qubit(i) in single operation)

        Teleport (qubit(i), Position)

        Position -=1

    endif
```

Figure 3 Algorithm for grid rearrangement.

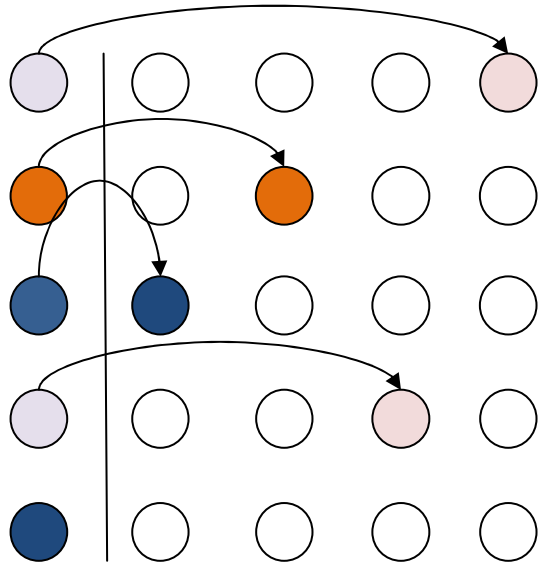

Figure 4 Applying algorithm for grid rearrangement in an example configuration. Qubits of the same color will partake in a single two-qubit operation. The leftmost column is the initial configuration, after the transposition the qubits are teleported to the right in the position indicated by the corresponding color.

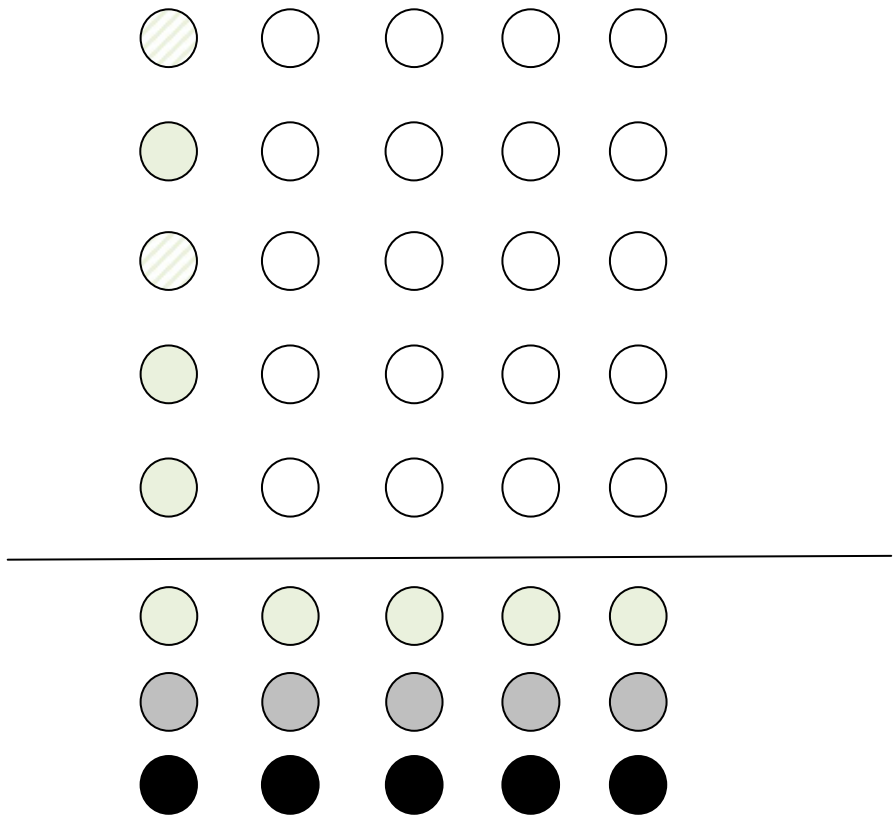

**Figure 5** Initial configuration of computational grid. The horizontal line separates the qubits partaking to the main computation from the ancilla qubits.

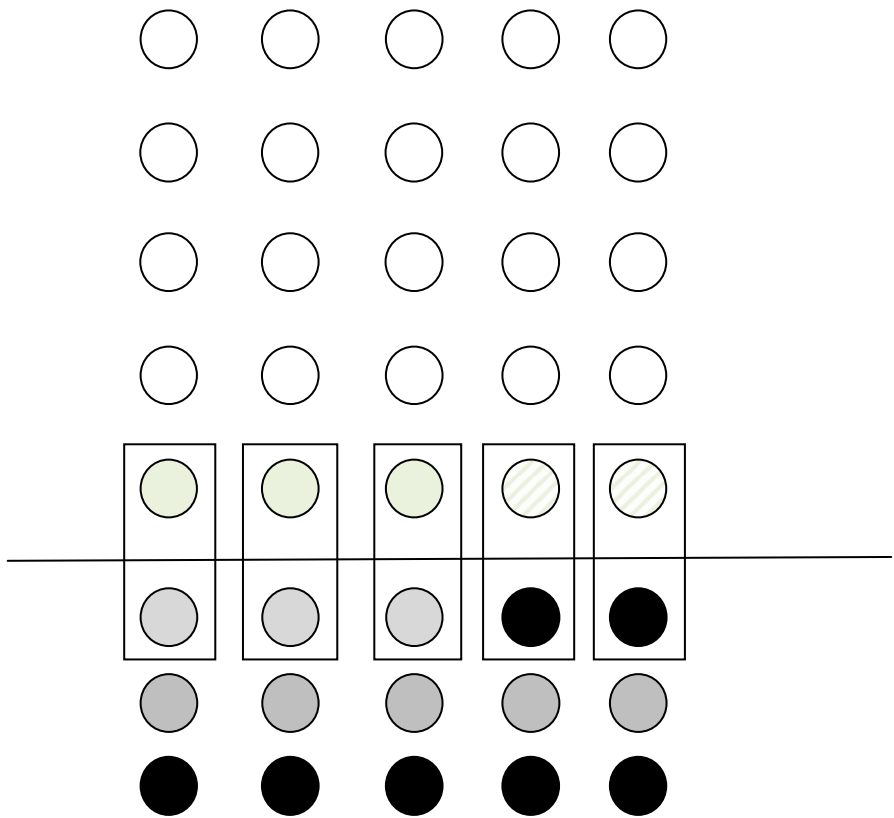

**Figure 6 Single qubit operation - Grid configuration**

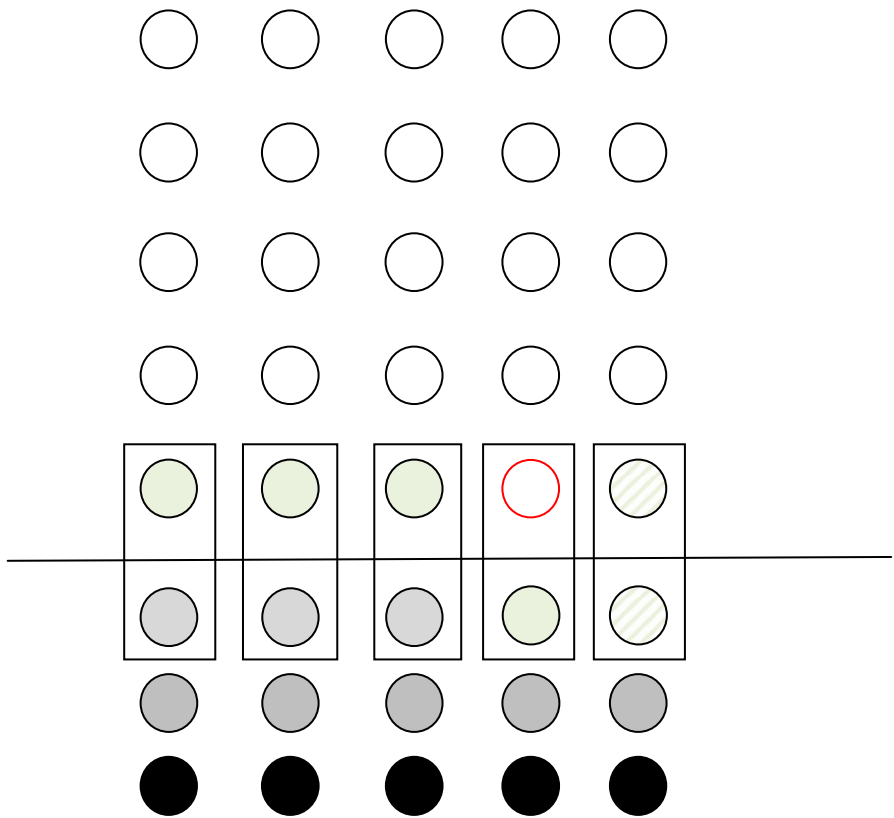

**Figure 7 Two qubit operation - Grid configuration**

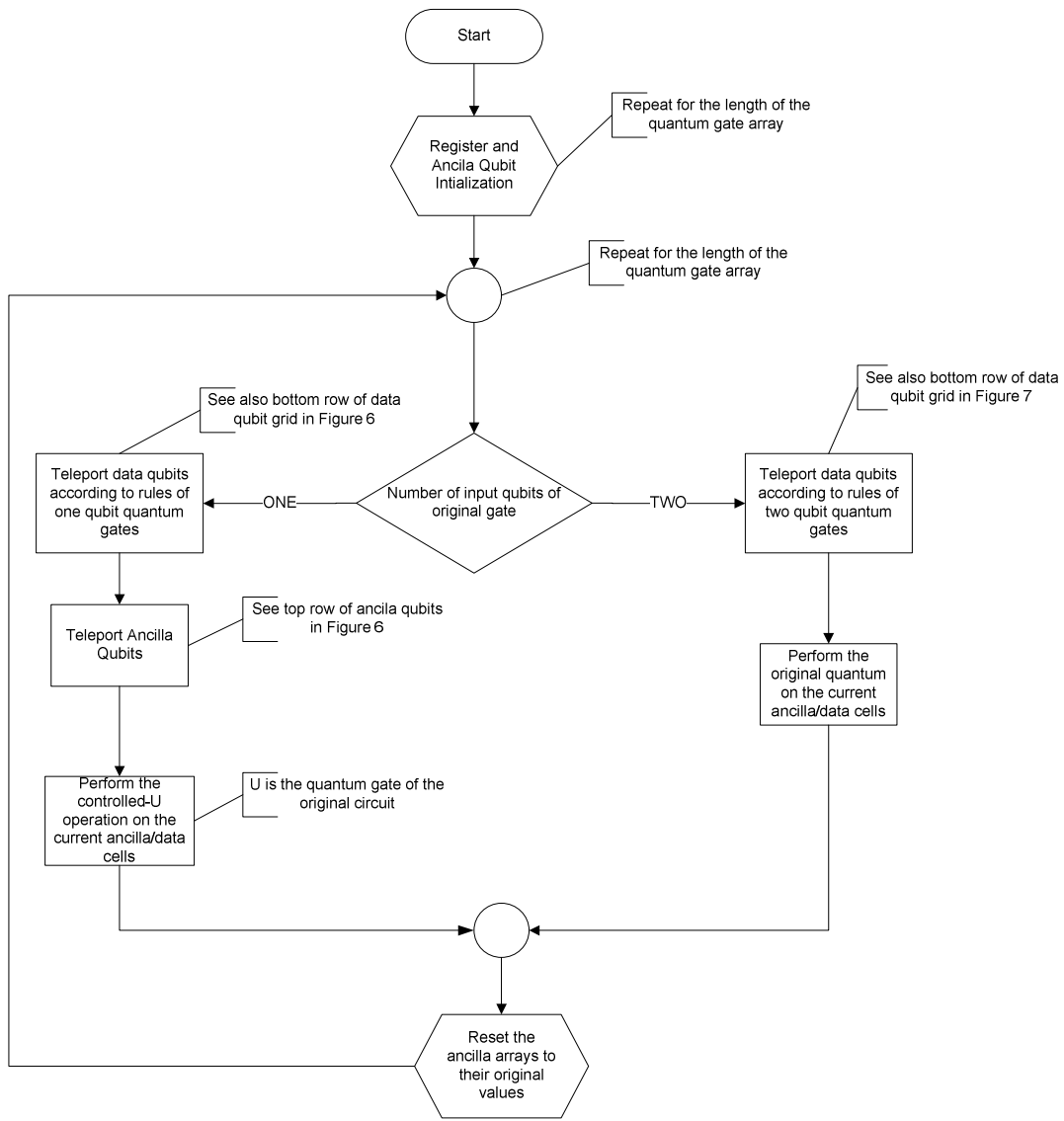

Figure 8 Flowchart of the generic algorithm. Main loop exit and algorithm termination are omitted

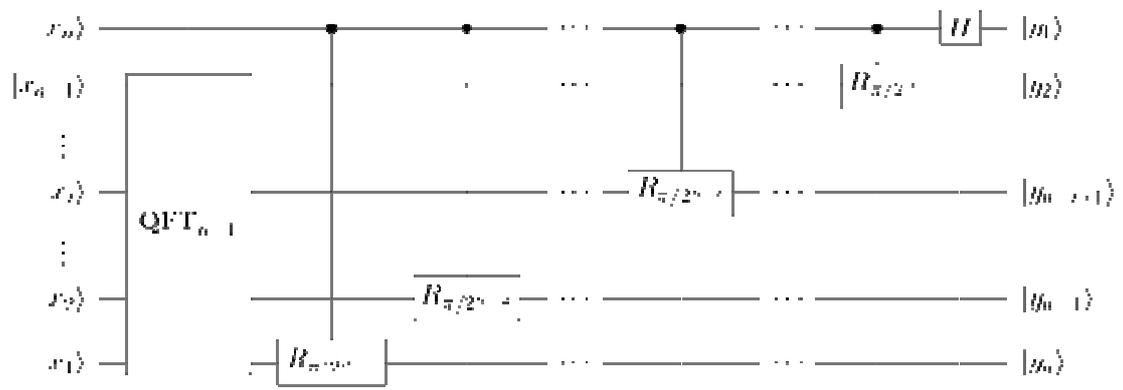

Figure 9 Quantum Fourier Transform

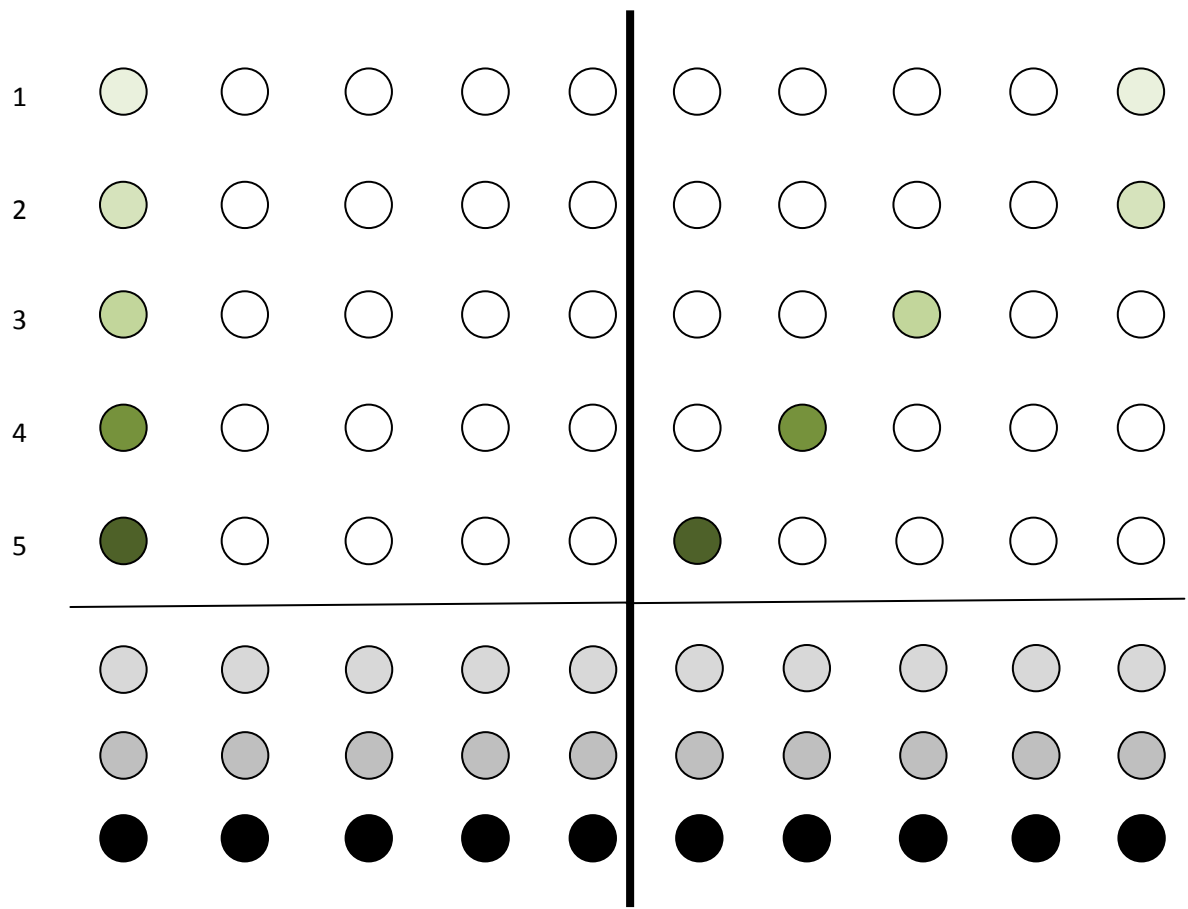

Figure 10: Controlled $R_{\pi/2}$. From initial configuration (left side) to the first horizontal teleportation. Qubits 1 and 2 will interact so they are teleported to the same column.

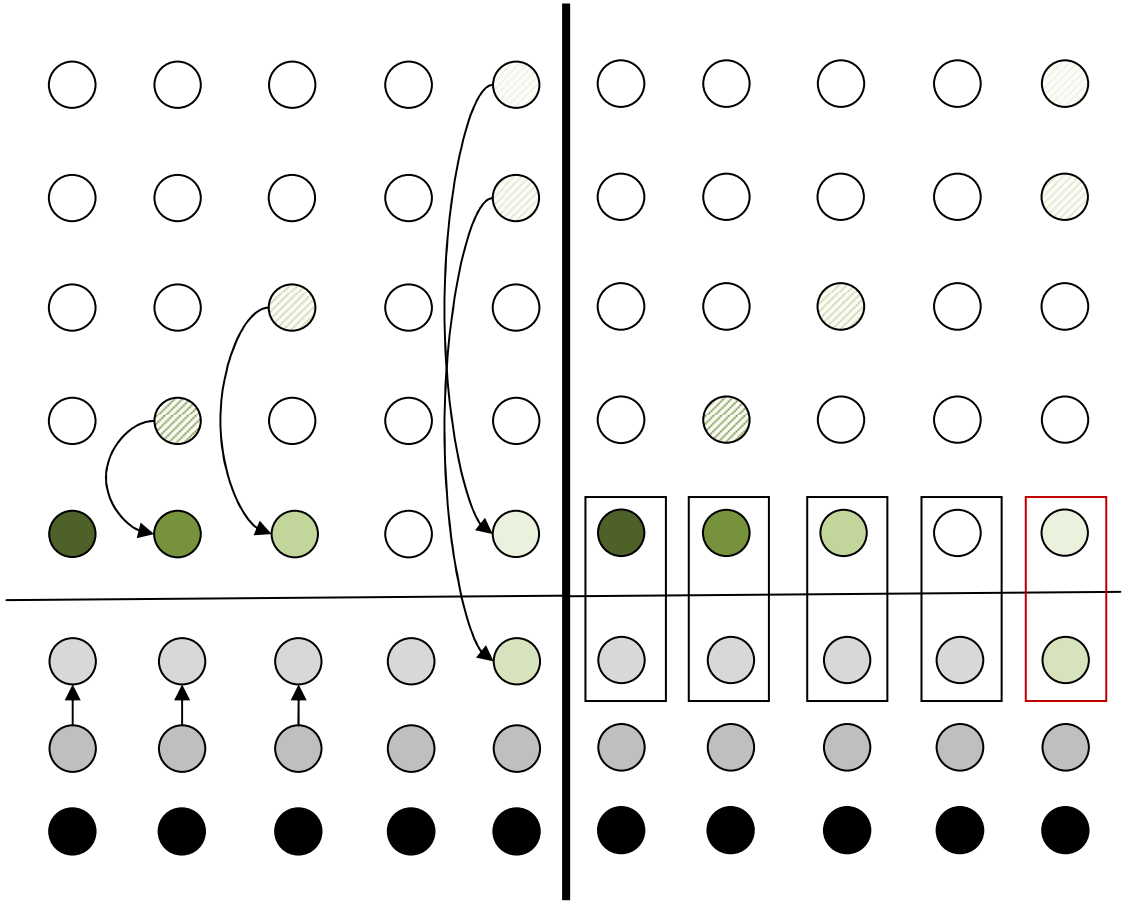

Figure 11: Controlled $R_{\pi/2}$. Vertical teleportation and applying the two-qubit quantum operation

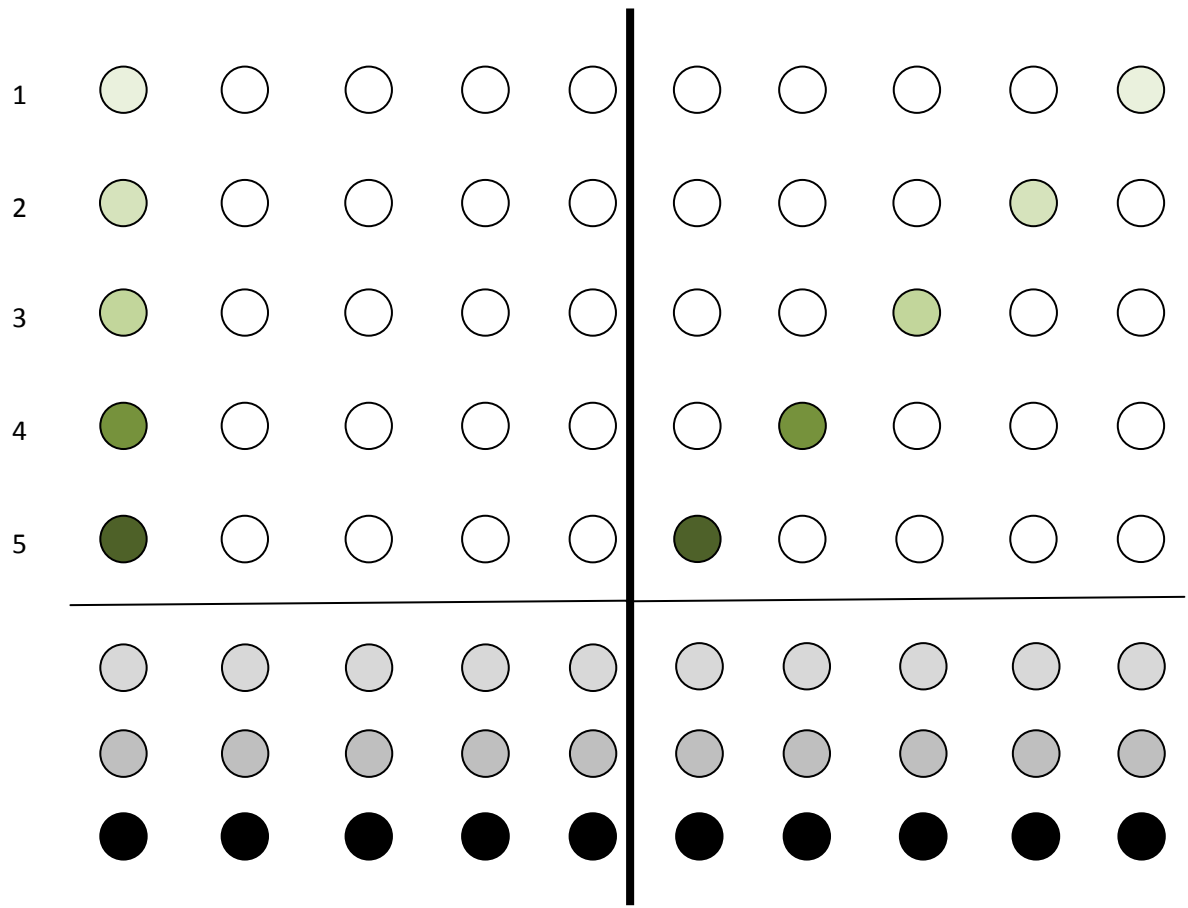

Figure 12: Hadamard Gate. From initial configuration to the first horizontal teleportation

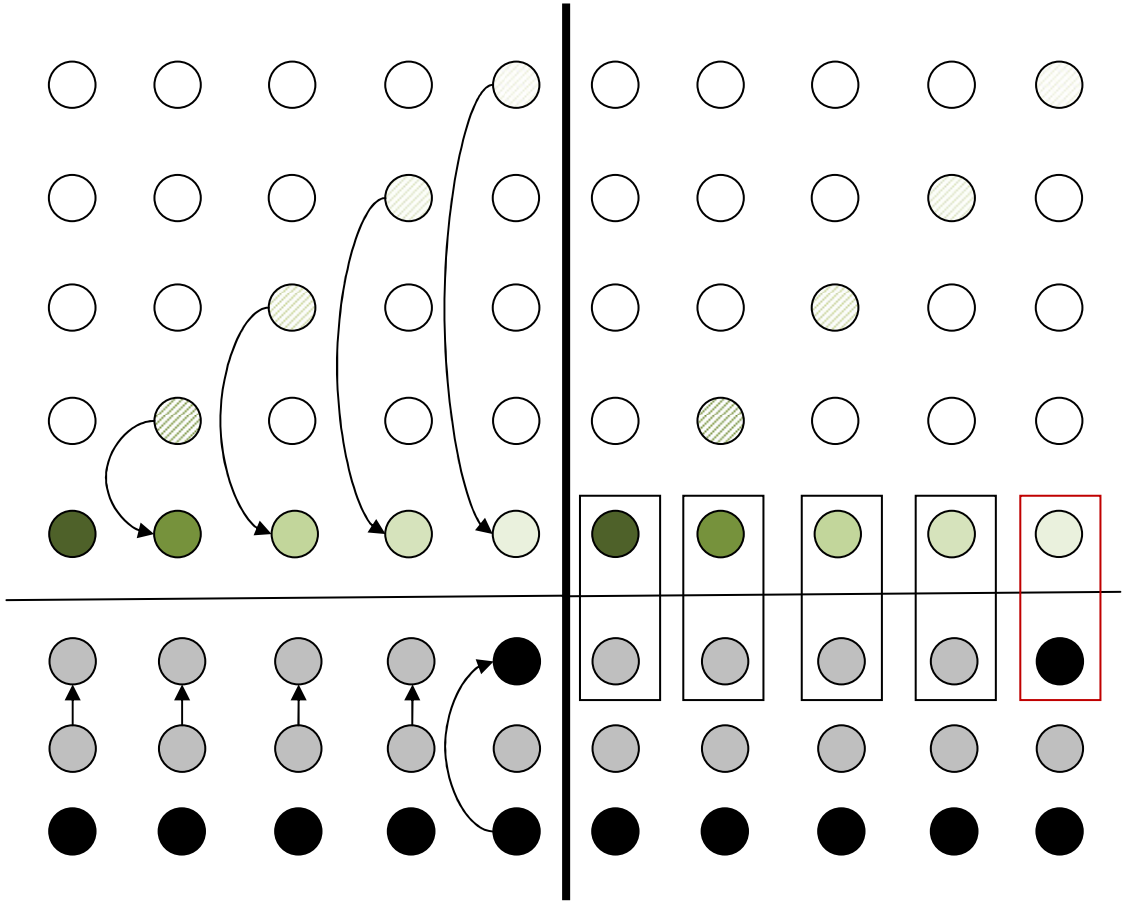

Figure 13: Hadamard gate. Vertical teleportation and applying the gate using control qubits in pre-prepared states.

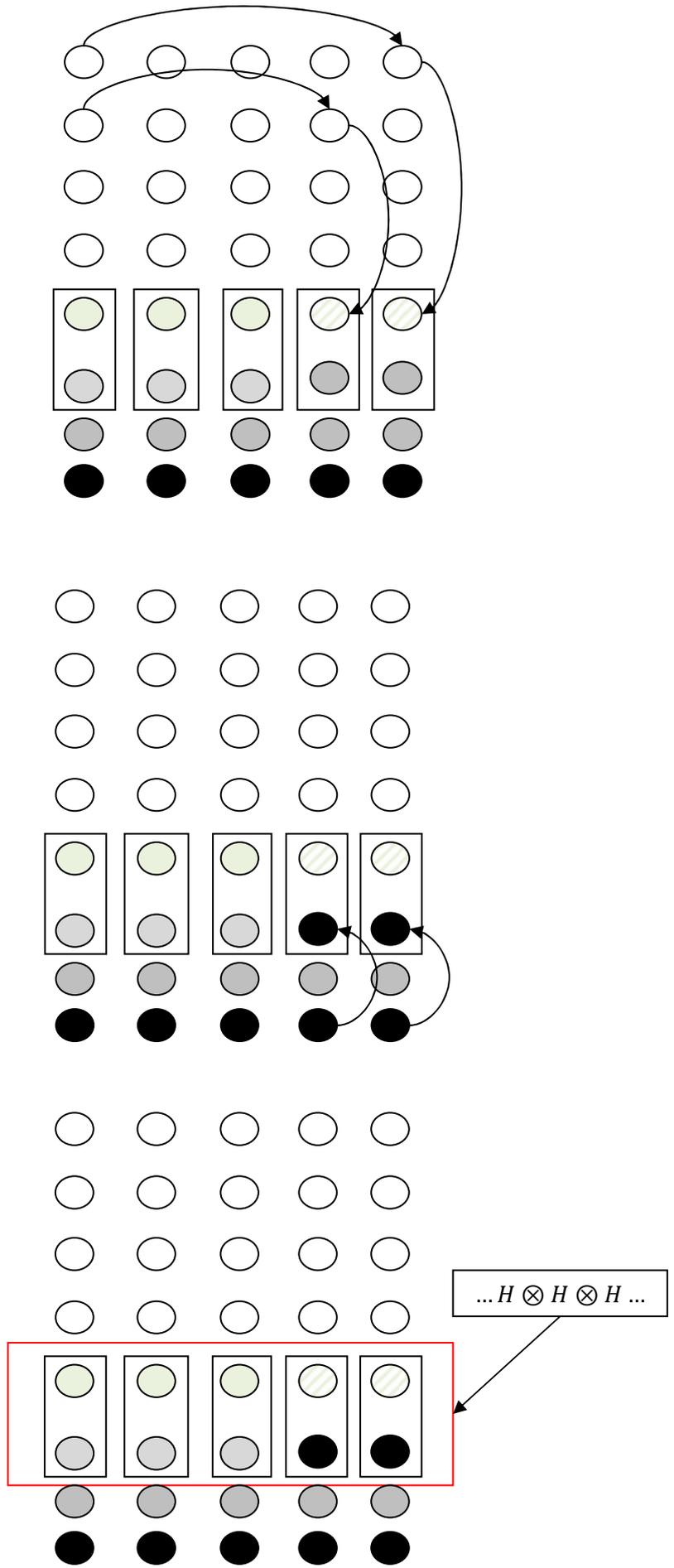

**Figure 14 Stages for the execution of the Hadamard gate. It is implied that the operation performed in the last stage is the controlled Hadamard operation.**

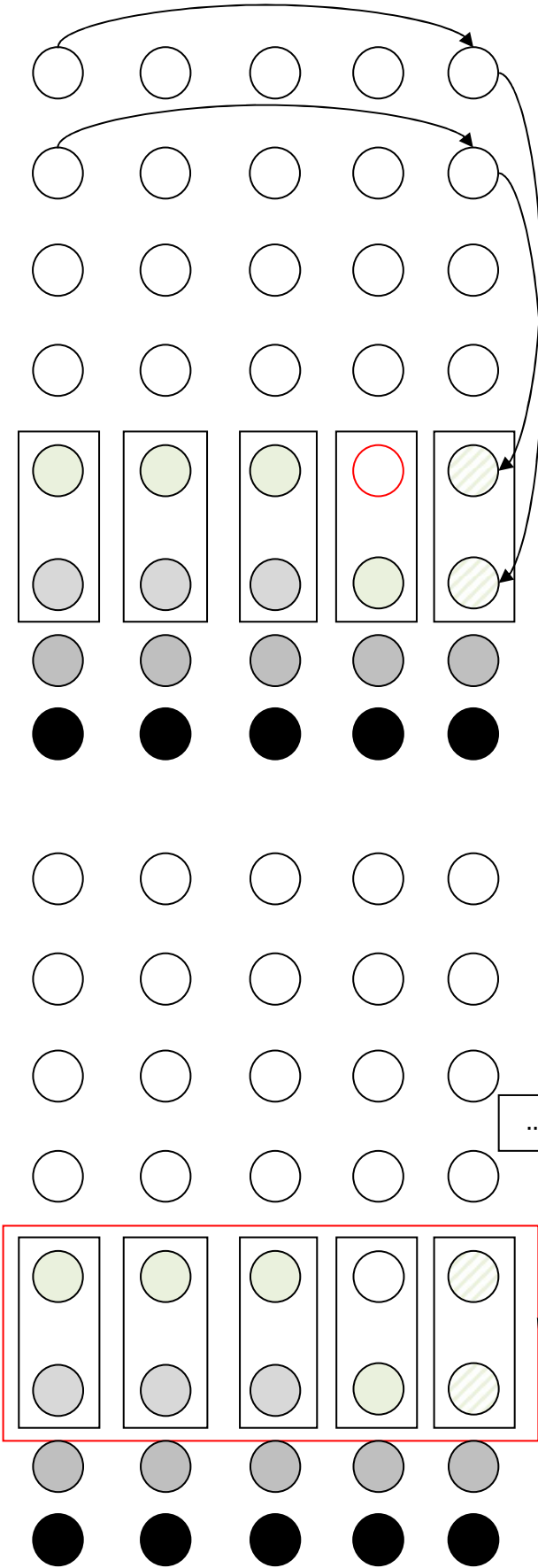

**Figure 15 Stages for the execution of the Controlled Phase operation**